\documentclass[aoas,seceqn,nameyear,rotating,MSNbibl,dvips]{arximspdf}

\doi{10.1214/10-AOAS368}
\volume{4}
\issue{4}
\pubyear{2010}
\firstpage{2126}
\lastpage{2149}

\begin{document}
\begin{frontmatter}

\title{Bayesian semiparametric inference for multivariate
doubly-interval-censored data}
\runtitle{Linear dependent Poisson--Dirichlet survival model}

\begin{aug}
\author[A]{\fnms{Alejandro} \snm{Jara}\corref{}\thanksref{t1}\ead[label=e1]{ajara@mat.puc.cl}
\ead[label=u1,url]{http://www.mat.puc.cl/\texttildelow ajara}},
\author[B]{\fnms{Emmanuel} \snm{Lesaffre}\ead[label=e2]{e.lesaffre@erasmusmc.nl}
\ead[label=u2,url]{http://www.erasmusmc.nl/biostatistiek/}},
\author[C]{\fnms{Maria} \snm{De Iorio}\ead[label=e3]{m.deiorio@imperial.ac.uk}
\ead[label=u3,url]{http://www1.imperial.ac.uk}} and
\author[D]{\fnms{Fernando} \snm{Quintana}\thanksref{t2}\ead[label=e4]{quintana@mat.puc.cl}
\ead[label=u4,url]{http://www.mat.puc.cl/\texttildelow quintana}}
\runauthor{Jara, Lesaffre, De Iorio and Quintana}
\thankstext{t1}{Supported in part by Fondecyt Grant 3095003.}
\thankstext{t2}{Supported in part by Fondecyt Grants 1060729 and
1100010, and Laboratorio de An\'alisis Estoc\'astico PBCT-ACT13.}
\affiliation{Pontificia Universidad Cat\'olica de Chile, %
Catholic University of Leuven  and %\thanksmark{m2},
Erasmus~University Rotterdam, %\thanksmark{m3}
Imperial College London %\thanksmark{m4}
and Pontificia~Universidad Cat\'olica de Chile}
\address[A]{A. Jara\\
Department of Statistics\\
Faculty of Mathematics\\
Pontificia Universidad Cat\'olica de Chile\\
Casilla 2, Correo 22\\
Chile\\
\printead{e1}\\
\printead{u1}}

\address[B]{E. Lesaffre\\
Department of Biostatistics\\
Erasmus Medical Centre\\
Erasmus University Rotterdam\\
3000 CA Rotterdam\\
The Netherlands \\
\printead{e2}\\
\printead{u2}}

\address[C]{M. De Iorio\\
Department of Epidemiology and\\
\quad Biostatistics\\
Imperial College London\\
W2 1PG\\
UK\\
\printead{e3}\\
\printead{u3}}

\address[D]{F. Quintana\\
Department of Statistics\\
Faculty of Mathematics \\
Pontificia Universidad Cat\'olica de Chile\\
Casilla 2, Correo 22\\
Chile\\
\printead{e4}\\
\printead{u4}}

\end{aug}

% HISTORY:
\received{\smonth{2} \syear{2009}}
\revised{\smonth{5} \syear{2010}}

% ABSTRACT
%
\begin{abstract}
Based on a data set obtained in a dental longitudinal study,
conducted in Flanders (Belgium), the joint time to caries
distribution of permanent first molars was modeled as a function of
covariates. This involves an analysis of multivariate
continuous doubly-interval-censored data since: (i) the emergence time
of a
tooth and the time it experiences caries were recorded yearly, and
(ii) events on teeth of the same child are dependent. To model the
joint distribution of the emergence times and the times to caries,
we propose a dependent Bayesian semiparametric model. A major feature
of the proposed approach is that survival curves can be estimated
without imposing assumptions such as proportional hazards, additive
hazards, proportional odds or accelerated failure time.
\end{abstract}

% KEYWORDS
%

\end{frontmatter}

%s1 ###
\section{Introduction}

The past three decades have witnessed a dramatic decline in the
prevalence of dental caries in children in countries of the Western
World [\citet{devosvano06}]. However, the disease has now become
concentrated in a small group of children, with the majority
unaffected; about 10--15\% of the children now experience 50\% of all
caries lesions and 25--30\% suffer 75\% of lesions [\citet{martomull96}; \citet{petebrat96}]. The most likely explanation for the
difference in oral health seems to be socio-economic environmental
factors and it occurs early in childhood [\citet{willvanomart05}].
Therefore, to improve dental health, early identification of groups at
a particular risk of developing caries becomes essential. In this paper
we present a Bayesian analysis of a longitudinal data set, gathered in
the Signal-Tandmobiel\tsup{\textregistered} study, to investigate the
relationship between some potential exposure variables and the
emergence and development of caries in permanent teeth.

%
%f1 ###
\begin{figure}[b]

\includegraphics{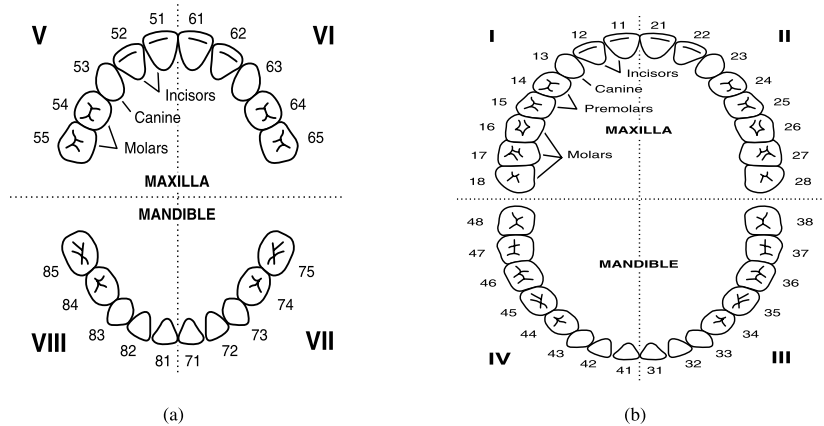}

%{
%}\\
%{
%}
\caption{European notation for the position of \textup{(a)}
deciduous (primary); and \textup{(b)} permanent teeth.
Maxilla $=$ upper jaw, mandible $=$ lower jaw. In \textup{(a)} the fifth and
the eight quadrants are at the right-hand side of the subject, and the
sixth and the seventh quadrants are to the left. In \textup{(b)} the first and
the fourth quadrants are at the
right-hand side of the subject, and the second and the third quadrants are
to the right.}\label{fig:teeths}
\end{figure}

The Signal-Tandmobiel\tsup{\textregistered} study is a 6-year longitudinal
oral health study involving children from Flanders
(Belgium) and conducted between 1996 and 2001. Dental data were\vadjust{\goodbreak}
collected on
gingival condition, dental trauma, tooth decay, presence of
restorations, missing teeth, stage of tooth eruption, orthodontic
treatment need, etc. Additionally, information on oral hygiene and
dietary behavior was collected from a questionnaire completed by the
parents. The children were examined
annually during their primary school time by one of sixteen trained
and half yearly calibrated dental examiners. More details on the
Signal-Tandmobiel\tsup{\textregistered} study can be found in Section~\ref{sec4.1} and in
\citet{vanobergenetal2000}. A primary objective of the
investigation is to assess the association of some covariates with
the emergence and development of caries in permanent teeth. In
particular, we are interested in studying the effect of the age at
start brushing (in years) and of deciduous second molars health status
[sound/affected; teeth 55, 65, 75, 85, respectively, see Figure~\ref{fig:teeths}a] on caries susceptibility of the adjacent permanent
first molars [teeth number 16, 26, 36, 46, see Figure~\ref
{fig:teeths}b]. Additionally, we considered the impact of gender
(girl/boy), presence of sealants in pits and fissures of the permanent
first molar (none/present), occlusal plaque accumulation on the
permanent first molar (none/in pits and fissures/on total surface) and
reported oral brushing habits (not daily/daily). Note that pits and
fissures sealing is a preventive action which is expected to protect
the tooth against caries development. The information on occlusal
plaque accumulation, presence of sealants in pits and fissures and
reported oral brushing habits was obtained at the examination where the
presence of the permanent first molar was first recorded.

\begin{figure}

\includegraphics{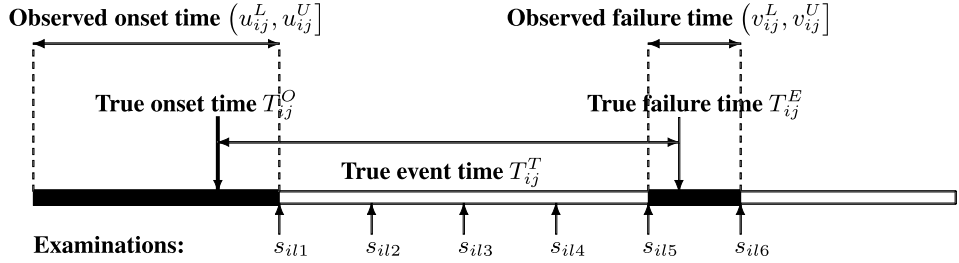}

%[ht]
%%
%
%
%
%
%35){\textbf{True failure time $T^E_{ij}$}}
%
%
%21){\line(1, 0){40}} \linethickness{0.4pt}
%
%
%
%u_{ij}^U ]$}}
%
%v_{ij}^U ]$}}
%
%%
\caption{An example of doubly interval censoring. A~scheme of
a~doubly-interval-censored observation obtained by performing
examinations to check the event status at times $s_{il1}, \ldots,
s_{il6}$. The onset time is left-censored at time $u_{i,l}^U =
s_{il1}$, that is, interval-censored in the interval $ ( u_{i,l}^L,
u_{i,l}^U  ] =  (0, s_{il1} ]$, the failure time is
interval-censored in the interval $ ( v_{i,l}^L, v_{i,l}^U
 ] =  ( s_{il5}, s_{il6}  ]$.}
\label{figDoublyCensoring}
\end{figure}

The response of interest is the time to caries development on the
permanent dentition which corresponds to the time from tooth emergence
to onset of caries. Due to the setup of the study (annual visits of
dentists), the onset time and the failure time could only be recorded
at regular intervals and observations on both events were, therefore,
interval-censored. A graphical illustration of a possible evolution of a
tooth is shown in Figure~\ref{figDoublyCensoring}. This type of data
structure, often referred to as doubly-interval-censored failure time
data, is common in medical research, especially in the context of the
analysis of acquired immunodeficiency syndrome (AIDS) incubation time,
the time between the human immunodeficiency virus infection and the
diagnosis of AIDS.
%
%f2 ###

Several approaches have been proposed over the past few years for
the analysis of doubly-interval-censored data. \citet{degruttolalagakos89} suggested a nonparametric maximum likelihood
(NPML) estimator of univariate survival functions. Alternative methods
were subsequently given by \citet{bacchettijewell91},
\citet{gomezlagakos94},
\citet{sun95} and \citet{gomezcalle99}.
\citet{kimdegruttolalagakos93} generalized the one-sample
estimation procedure of \citet{degruttolalagakos89} to a Cox
proportional hazards (PH) model. Their method, however, needs to
discretize the data. Cox regression with the onset time
interval-censored and the event time right-censored has been
considered by \citet{gogginsfinkelsteinzaslavsky99},
\citet{sunliaopagano95} and \citet{pan2001}. To simplify the
analysis, all of these methods make a rather unrealistic independence
assumption between the onset and time-to-event variables [see,
e.g., \citet{sunlimzhao04}].

For the analysis of multivariate doubly-interval-censored survival
data, frailty models were discussed in
\citet{komareklesaffreharkanendeclerckvirtanen2005} and
\citet{komareklesaffre2008} considering versions of the Cox PH and
accelerated failure time (AFT) models, respectively. In the latter
case, each distributional part is specified in a flexible way as a
penalized Gaussian mixture with an overspecified number of mixture
components and under the assumption of independence between the onset
and time-to-event variables. These models provide useful summary
information in the
absence of estimates of a baseline survival distribution and may be
formulated in a parametric or semi-parametric fashion. However, under
these models the
regression coefficients describe changes in individual responses due to changes
in covariates, they induce a particular association structure for the\vadjust{\goodbreak}
clustered variables, and rely heavily on the (conditional or
subject-specific) assumptions
of PH or AFT in the relationship between the covariates and the
survival times. While the PH model assumes the covariates act
multiplicatively on a baseline hazard function, the AFT model
assumes that covariates act multiplicatively on arguments of the
baseline survival function. Although other type of models, such as
additive hazards (AH) or proportional odds (PO), could be considered
in a frailty model context, all these assumptions may be considered too
strong in
many practical applications. For instance, under these models survival
curves from different covariate groups cannot cross which can be
unrealistic in some applications [see \citet{deioriorjohnsonmuellerrosnermaceachern2009}]. This issue is
particularly relevant for doubly-interval-censored data where the
degree of available information to perform diagnostic techniques is
rather reduced due to the censoring mechanism.

In this paper we discuss a Bayesian semiparametric approach for the
analysis of multivariate doubly-interval-censored data where the
dependence across sub-populations, defined by different combinations
of the available covariates, is introduced without assuming
independence between the onset and time-to-event variables, without
requiring data discretization, and any of
the commonly used \mbox{assumptions} for the inclusion of covariates in
survival models. We extend recent developments
on dependent nonparametric priors, initially proposed by
\citeauthor{maceachern99} (\citeyear{maceachern99},
\citeyear{maceachern2000}), to provide a framework for modeling
multivariate doubly-interval-censored data where the resulting
survival curves have a marginal (or population level) interpretation
and are not subject-specific.
It must be pointed out that the dental data has been analyzed before.
However, the previous
approaches were deficient in that either the doubly-interval-censored
nature was not taken into account [\citet{leroybogaertslesaffredeclerck2005a}] or restrictive in the
sense that the focus was on conditional interpretation of the
effects of the covariates via frailty models and relying on the AFT or
PH assumption
[\citet{komareklesaffreharkanendeclerckvirtanen2005}; \citet{komareklesaffre2008}].
Overcoming these problems largely motivates the developments
presented in this paper.

The rest of the paper is organized as follows. In
Section~\ref{LDPD_secBSPM} we introduce the proposed model, which is
based on the two parameter Poisson--Dirichlet process, and discuss its
main properties. Section~\ref{LDPD_sec3} presents the analysis of
simulated data which illustrate the main advantage of the proposed model.
Section~\ref{LDPD_sec4} describes the analysis of the Signal-Tandmobiel\tsup{\textregistered} study. A final discussion section concludes
the article.

%s2 ###
\section{The model}\label{LDPD_secBSPM}

%s2.1 ###
\subsection{Survival regression framework}\label{LDPD_sec2.1}

Let $T^O_{ij}$ and $T^E_{ij}$, $i=1,\ldots,m$, $j=1,\ldots,n$, be
continuous random variables defined on $[0, \infty)$
denoting the true chronological onset and event times for the $j$th
measurement of the $i$th experimental unit, respectively, and let
$T^T_{ij} = T^E_{ij}- T^O_{ij}$ be the true time-to-event.
For example, in our case $T^T_{ij}$ is the true time to caries for the
$j$th tooth of the $i$th child, with $T^O_{ij}$ denoting the true
emergence time and $T^E_{ij}$ the age of caries development. Assume that
for each of the $m$ experimental units we record the $p$-dimensional
and $q$-dimensional covariate
vectors $\mathbf{x}^O_{ij} \in\mathcal{X}^O \subset
\mathbb{R}^p$ and $\mathbf{x}^T_{ij} \in\mathcal{X}^T \subset
\mathbb{R}^q$ associated to the onset time $T^O_{ij}$ and to the
time-to-event $T^T_{ij}$, respectively. Let
$\mathbf{T}^O_i= (T^O_{i1},\ldots,T^O_{in} )'$,
$\mathbf{T}^E_i= (T^E_{i1},\ldots,T^E_{in} )'$,
$\mathbf{T}^T_i= (T^T_{i1},\ldots,T^T_{in} )'$,
$\mathbf{T}_i= (\mathbf{T}^{O'}_i,\mathbf{T}^{T'}_i )'$,
$\mathbf{X}^O_i=\operatorname{diag} (\mathbf{x}^{O'}_{i1},\ldots
,\mathbf{x}^{O'}_{in} )$,
$\mathbf{X}^T_i=\operatorname{diag} (\mathbf{x}^{T'}_{i1},\break\ldots
,\mathbf{x}^{T'}_{in} )$
and
$\mathbf{X}_i=\operatorname{diag} (\mathbf{X}^{O}_{i},\mathbf{X}^{T}_{in} )$,
$i=1,\ldots,m$.

In order to model the joint distribution of the true chronological
onset times and true time-to-events $\mathbf{T}_i$ as a function
of covariates, $\mathbf{X}_i$, we consider a mixture model.
Specifically, we
assume $\mathbf{T}_i\vert\mathbf{X}_i\stackrel{\mathrm{i.n.d.}}{\sim}
f_{\mathbf{X}_i}$, $i=1,\ldots,m$, with
%
%e2.1 ###
\begin{eqnarray}\label{LDPD_mixture}
f_{\mathbf{X}_i}  (\cdot\vert\bolds{\Sigma},
G_{\mathbf{X}_i} )=\int k_{2 n} (\cdot\vert
\bolds{\mu}, \bolds{\Sigma}  )\, d G_{\mathbf{X}_i}
 (\bolds{\mu}  ),
\end{eqnarray}
where $k_{2 n} (\cdot\vert\bolds{\mu},
\bolds{\Sigma} )$ denotes a $2n$-variate density on
$\mathbb{R}_+^{2n}$ with location $\bolds{\mu}$ and unstructured scale
matrix $\bolds{\Sigma}$ taking into account the association among
variables of the same experimental unit, respectively, and where
the mixing distributions $G_{\mathbf{X}_1},\ldots,G_{\mathbf{X}_m} \in \{G_{\mathbf{X}}\dvtx  \mathbf{X} \in\mathcal{X}
 \}$ are dependent probability measures. The set of dependent
probability measures $ \{G_{\mathbf{X}}\dvtx  \mathbf{X} \in
\mathcal{X}  \}$ is defined in the complete space of the
predictors $ \mathcal{X}$ and the degree of dependence among the
elements is governed by the value of the covariates $\mathbf{X}$.
If $G_{\mathbf{X}}$
were indexed by a finite-dimensional vector of hyper-parameters, for
example, normal moments, then the model would reduce to a
traditional parametric hierarchical model. In contrast, in a
nonparametric Bayesian approach, every element in the set $\{
G_{\mathbf{X}}\dvtx  \mathbf{X} \in\mathcal{X}\}$ is a random
probability measure and an appropriate prior
probability model $F$ for the complete set of unknown distributions
indexed by the
set of covariates $\{G_{\mathbf{X}}\dvtx  \mathbf{X} \in\mathcal{X}\}$ is specified. In other words, $F$ is a distribution over
related probability distributions
%
%e2.2 ###
\begin{eqnarray}\label{LDPD_rrpm}
 \{G_{\mathbf{X}}\dvtx  \mathbf{X} \in\mathcal{X}  \}
\vert F \sim F.
\end{eqnarray}
Here we focus on the class of discrete random probability measures
that can be represented as
%
%e2.3 ###
\begin{eqnarray}\label{LDPD_dldm}
G_{\mathbf{X}}(B)=\sum_{l=1}^{\infty} \omega_l
\delta_{\bolds{\theta}(\mathbf{X})_l}(B),
\end{eqnarray}
where $B$ is a measurable set, $\omega_1, \omega_2, \ldots$ are
random weights satisfying $0 \leq\omega_l \leq1$ and
$P(\sum_{l=1}^{\infty}$ $ \omega_l$ $=1)=1$, and where
$\delta_{\bolds{\theta}(\mathbf{X})_l}(\cdot)$ denotes a
Dirac measure at the random locations
$\bolds{\theta}(\mathbf{X})_1,\bolds{\theta}(\mathbf{X})_2,\ldots,$
which are assumed to be independent of the
$\{\omega_l\}_{l>1}$ collection. We discuss
specific choices for the random probability measure~$F$
in~(\ref{LDPD_rrpm}) in the next sections. To better explain our
proposal, we start with a review of the construction of priors over
related distributions.

%s2.2 ###
\subsection{Priors over related distributions}\label{LDPD_sec2.2}

The problem of defining priors over related random probability
distributions has received increasing attention over the past few
years. \citeauthor{maceachern2000} (\citeyear{maceachern99},
\citeyear{maceachern2000}) proposes the dependent Dirichlet Process
(DDP) as an approach to define a prior model for an uncountable set
of random measures indexed by a single continuous covariate, say,
$x$ $\{G_{x}\dvtx  x \in\mathcal{X} \subset\mathbb{R}\}$. The key idea
behind the DDP is to create an uncountable set of Dirichlet
Processes (DP) [\citet{ferguson73}] and to introduce dependence by
modifying the \citeauthor{sethuraman94}'s (\citeyear{sethuraman94}) stick-breaking representation
of each element in the set. If $G$ follows a DP prior with precision
parameter $M$ and base measure $G_0$, denoted by $G \sim \mathit{DP}(M G_0)$,
then the stick-breaking representation of $G$ is
%
%e2.4 ###
\begin{eqnarray}\label{stickbreaking}
G(B)=\sum_{l=1}^\infty\omega_l \delta_{\theta_l}(B),
\end{eqnarray}
where $\theta_l \vert G_0 \stackrel{\mathrm{i.i.d.}}{\sim} G_0$ and $\omega_l=V_l
\prod_{j<l}(1-V_j)$, with $V_l \vert M \stackrel{\mathrm{i.i.d.}}{\sim}
\operatorname{Beta}(1,M)$. \citeauthor{maceachern2000}
(\citeyear{maceachern99}, \citeyear{maceachern2000}) generalizes
(\ref{stickbreaking}) by assuming the point masses $\theta(x)_l$,
$l=1,\ldots,$ to be dependent across different levels of $x$, but
independent across $l$. This approach has been successfully applied
to ANOVA [\citet{deioriormuellerrosnermaceachern2004}], survival
[\citet{deioriorjohnsonmuellerrosnermaceachern2009}], spatial
modeling [\citet{gelfandkottasmaceachern2005}], functional data
[\citet{dunsonherring2006}], time series
[\citet{carondavydoucetduflosvanheeghe2007}] and discriminant
analysis [\citet{delacruzquintanamueller2007}]. Motivated by
regression problems with continuous predictors,
\citet{griffinsteel2006} and \citet{duanguidanigelfand2007}
developed models where the dependence is introduced by making the
weights dependent on covariates.

Alternatives to these approaches include incorporating dependency by
means of weighted mixtures of independent random measures
[\citet{mullerquintanarosner2004}; \citet{dunsonpark2008}]. This approach
was originally proposed by \citet{mullerquintanarosner2004},
motivated for the problem of borrowing strength across related
submodels. For regression problems with continuous
predictors, \citet{dunsonpark2008} proposed a countable mixture
where the weights depend on the covariates through the introduction
of a bounded kernel function in the stick-breaking construction of
the weights. The latter approach requires the choice of a metric for the
covariate values and, therefore, is not naturally extended to
include factors and continuous predictors jointly in the model.

We build our proposal on the construction introduced in
\citet{deioriormuellerrosnermaceachern2004} and
\citet{deioriorjohnsonmuellerrosnermaceachern2009} because it is a
natural approach to introduce dependence on both factors and
continuous covariates which are commonly of interest in survival
models. We consider the class of discrete Linear Dependent (LD)
models defined as follows. For any given value of the covariates
$\mathbf{X} \in\mathcal{X}$, in the notation of our motivating problem,
the $2n$-dimensional atoms in the
mixing distribution $
G_{\mathbf{X}}(\cdot)=\sum_{l=1}^{\infty} \omega_l
\delta_{\bolds{\theta}(\mathbf{X})_l}(\cdot)
$ follow linear (in the parameters) models
$\bolds{\theta}(\mathbf{X})_l =
\mathbf{X}\bolds{\beta}_{l}$, where the
$\bolds{\beta}_{l}$'s represent $n(p+q)$-dimensional vectors
of regression coefficients. Therefore, in the dependent mixture model
given by expression (\ref{LDPD_mixture}), $P(\bolds{\mu}=\bolds{\theta}(\mathbf{X})_l=\mathbf{X}\bolds{\beta}_{l})
=\omega_l$ and the dependence is introduced in the
point mass locations $ \bolds{\theta}(\mathbf{X})_l$ through a
linear model, where the regression coefficients $\bolds{\beta}_{l}$ are i.i.d. random vectors from a distribution $G_0$,
$\bolds{\beta}_{l} \stackrel{\mathrm{i.i.d.}}{\sim}G_0$.
For simplicity of explanation, consider the case of $n=1$ and an ANCOVA
type of design matrix
\begin{eqnarray}\nonumber
\mathbf{X}=
\pmatrix{
1 & V & 0 & 0 & 0\cr
0 & 0 & 1 & V & Z
}
,
\end{eqnarray}
where $V$ is an indicator variable and $Z$ is continuous. For example,
$V$ could be the gender indicator and $Z$ the age at start brushing. In
the LD model the dependence across the random distributions is achieved
by imposing a linear model on the point masses
\begin{eqnarray}\nonumber
\bolds{\theta}(\mathbf{X})_l =\mathbf{X}\bolds{\beta}_{l}=
\pmatrix{
\beta_{1l} + \beta_{2l} V\cr
\beta_{3l} + \beta_{4l} V + \beta_{5l} Z
}
.
\end{eqnarray}
As in a standard linear model, $\beta_{1l}$ and $\beta_{3l}$ can be
interpreted as intercepts for the point masses associated to the onset
time and to the time-to-event, respectively, while $\beta_{2l}$ and
$\beta_{4l}$ are the main effects of gender for the onset and
time-to-event, respectively, and $\beta_{5l}$ can be interpreted as a
slope coefficient associated to the age at start brushing for the
time-to-event. Note that the linear specification is highly flexible
and can include standard nonlinear transformations
of the continuous predictors, for example, additive models based on
B-splines [see, e.g., \citet{langbrezger2004}], as well as linear
forms in the continuous predictors themselves.

%s2.3 ###
\subsection{The proposal}\label{LDPD_sec2.3}

In this paper we extend the DDP framework to a construction that is
based on
the general class of Poisson--Dirichlet (PD) processes [see, e.g., \citet{pitman96} and \citet{pitmanyor97}]. The PD processes belong to the
class of species sampling models [see, e.g., \citet{pitman96}] and
admit the DP prior as an important special case.
The PD process can also be defined as in expression (4), where the
random weights $\omega_l$ are independent for the $\theta_l$'s and the
$\theta_l$ are i.i.d. from a distribution $G_0$. The weights still
admit a stick-breaking representation $\omega_l=V_l\prod_{j<l}
(1-V_j)$, but in this case $V_j \stackrel{\mathrm{i.n.d.}}{\sim}\operatorname{Beta}(1-a,b+ja)$, where either $a=-\kappa<0$ and $b=\varsigma\kappa$,
for some
$\kappa>0$ and $\varsigma=2,3,\ldots,$ or $0\leq a <1$ and $b>-a$. We restrict
our attention to the parameter space $\mathcal{A}=\{(a,b)\in\mathbb{R}^2\dvtx  0\leq a <1,b>-a\}$ because
this is large enough to include two important special cases. When
$a=0$ and $b=M$, Ferguson's $\mathit{DP}(MG_0)$ follows. When
$a=\gamma$, $0<\gamma<1$, and $b=0$, the $\mathit{PD}(\gamma,0)$ yields a
measure whose random weights are based on a stable law with index
$\gamma$. The DP and stable law are key processes because they
represent the canonical measures of the PD process
[\citet{pitmanyor97}].

It is now straightforward to extend the Linear Dependent framework to
the PD process assuming a linear model for the atoms of the process. In
this way we can define
a model for related probability distributions of the form
%
%e2.5 ###
\begin{eqnarray}\label{LDPD_ldpy}
 \{G_{\mathbf{X}}\dvtx  \mathbf{X} \in\mathcal{X}  \}
\vert a,b,G_0 \sim \mathit{LDPD}(a,b,G_0),
\end{eqnarray}
where $\mathit{LDPD}(a,b,G_0)$ refers to a Linear Dependent PD prior, with
parameters $a$, $b$, and $G_0$. An appealing property of the LDPD
survival model given by
expressions~(\ref{LDPD_mixture}) and~(\ref{LDPD_ldpy}) is that it
can be understood on the basis of an equivalent model reformulation
as a mixture of multivariate AFT regression models. Given a particular
matrix of covariates $\mathbf{X} \in\mathcal{X}$, the
vector of kernel locations $\bolds{\mu}$ in the
mixture model (\ref{LDPD_mixture}) takes the value $\mathbf{X}
\bolds{\beta}$, where the mixture is defined with respect to the
regression coefficients
$\bolds{\beta}$. In other words, the model can be alternatively
formulated by defining the
mixture of multivariate regression models,
%
%e2.6 ###
\begin{eqnarray}\label{LDPD_mixtureLD}
f_{\mathbf{X}}  (\cdot\vert\bolds{\Sigma},
G )=\int k_{2 n} (\cdot\vert
\mathbf{X}\bolds{\beta}, \bolds{\Sigma}  )\, dG
 (\bolds{\beta} )
\end{eqnarray}
for all $\mathbf{X} \in\mathcal{X}$, and
%
%e2.7 ###
\begin{eqnarray}
G\vert a,b,G_0 \sim \mathit{PD}(a,b,G_0).
\end{eqnarray}
The
discrete nature of the PD realizations leads to their well-known
clustering properties. The choice of parameters $a$ and $b$ in the PD
process controls the
clustering structure [\citet{lijoimenapruenster2007a}]. Given $m$
observations, when $a=0$ (i.e., a DP) the number of
clusters $n^*(m)$ is a sum of independent indicator variables, which
implies $n^*(m)/\log m \rightarrow b$ almost surely and $n^*(m)$ is
asymptotically normal [\citet{korwardhollander73}]. Under the model
with $0< a <1$ and $b>-a$ the sequence $\{n^*(m)\}$ is an
inhomogeneous Markov chain such that $n^*(m)/ m^a \rightarrow S$
almost surely, for a random variable $S$ with a continuous density
on $(0,\infty)$ depending on $(a,b)$ [\citet{pitmanyor97}]. The
asymptotic behavior of the distribution of the number of clusters
indicates that a general PD model increases as $m^a$ which is much
faster than the logarithmic rate of the DP model. In general, values of
$a$ close to 1 favor the
generation of a larger number of clusters.

Besides the clustering structure implied by the extra $a$ parameter in
the PD process, its role can be also understood when the distribution
of PD realizations is applied to a partition of the space of interest.
In particular, for measurable sets $B$, $B_1$ and $B_2$, with $B_1\cap
B_2=\varnothing$, it follows that [\citet{carlton99}]
%
%e2.8 ###
\begin{eqnarray}
\operatorname{Var} (G(B) )=G_0(B)  \bigl(1-G_0(B)  \bigr)  \biggl( \frac{1-a}{b+1}  \biggr)
\end{eqnarray}
and
%
%e2.9 ###
\begin{eqnarray}
\operatorname{Cov} (G(B_1),G(B_2) )=-G_0(B_1) G_0(B_2)
\biggl(\frac{1-a}{b+1} \biggr).
\end{eqnarray}
Therefore, the extra $a$ parameter controls the variability and
covariance of disjoint sets of the PD realizations. When $a \rightarrow
1$, $G$ is highly concentrated around $G_0$ and the covariance between
disjoint sets is small.
When $a=0$ we recover the corresponding expressions for the DP. Note
that the correlation between $G(B_1)$ and $G(B_2)$ does not depend on
the parameter $(a,b)$ and, therefore, is the same as the one arising
from the DP model.

To date, most practical implementations of PD processes have
considered the parameters $a$ and $b$ as fixed at user-specified
values [see, e.g., \citet{ishwaranjames2001}], fixed at empirical
Bayes estimates [see, e.g., \citet{lijoimenapruenster2007c}], or
explored the effect of different combinations of fixed values for
these parameters on the inferences [see, e.g., \citet{navarretequintanamuller2008}].
\citet{lijoimenapruenster2008}, on the other hand, proposed
independent discrete uniform priors with support points $\{0.01,
0.02,\ldots, 0.99\}$ and $\{0, 1,\ldots, 2000\}$ for $a$ and $b$,
respectively. Here we allow $a$ and $b$ to be random, having
continuous random probability distributions supported on the
restricted parameter space under consideration. Moreover, we allow
$a$ to be zero with positive probability in order to test whether
the data arose from LDDP versus a more general LDPD process using a
Bayes factor. This additional flexibility can be
incorporated at essentially no additional computational cost.

%s2.4 ###
\subsection{The hierarchical representation}\label{LDPD_sec2.4}

So far, we have focused on modeling the joint distribution of the
survival times of interest, namely, the true chronological onset
times $T^O_{ij}$ and true times-to-event $T^T_{ij}$. However, in our
setting the observed data are given by the events $\{T^O_{ij} \in
(u^L_{ij},u^U_{ij}]\dvtx  i=1,\ldots,m, j=1,\ldots,n \}$, and $
\{T^E_{ij} \in$ $(v^L_{ij},v^U_{ij}]\dvtx  i=1,\ldots,m, j=1,\ldots,n
\}$, where $u^L_{ij}$ and $v^L_{ij}$, and $u^U_{ij}$ and $v^U_{ij}$,
represent the lower
and upper limits of the intervals where the chronological onset,
$T^O_{ij}$, and event time, $T^E_{ij}$, for observation $j$ from
experimental unit $i$ were observed, respectively. Under the assumption
of noninformative censoring, we define a
model for the events $\mathbf{A}^O_i=
\{T^O_{ij} \in(u^L_{ij},u^U_{ij}]\dvtx  j=1,\ldots,n  \}$ and $
\mathbf{A}^E_i= \{T^E_{ij} \in(v^L_{ij},v^U_{ij}]\dvtx
j=1,\ldots,n  \}$, by introducing latent vectors
$\mathbf{T}_i^O$ and $\mathbf{T}_i^E$. We assume
%
%e2.10 ###
\begin{eqnarray}\label{LDPD_hier.latent}
(\mathbf{T}_i^O,\mathbf{T}_i^E
 )\vert h_{\mathbf{X}_i} \stackrel{\mathrm{i.n.d.}}{\sim}
h_{\mathbf{X}_i},
\end{eqnarray}
with
$h_{\mathbf{X}_i} (\mathbf{T}_i^O,\mathbf{T}_i^E
\vert\bolds{\Sigma}, G )\equiv
f_{\mathbf{X}_i} (\mathbf{T}_i^O,\mathbf{T}_i^E-\mathbf{T}_i^O
\vert\bolds{\Sigma}, G )$ and where
$f_{\mathbf{X}_i}(\cdot\vert \bolds{\Sigma},$ $G)$ is
defined as in~(\ref{LDPD_mixtureLD}). Notice that a choice of the
continuous kernel $k$ defines the model. A multivariate log-normal
distribution is convenient for practical reasons. Let
$\mathbf{z}_i=(\log T^O_{i1},\ldots, \log T^O_{in},\log
T^T_{i1},\ldots,$ $\log T^T_{in})'$ denote the logarithmic
transformation of the true chronological onset times and true
times-to-event such that
%
%e2.11 ###
\begin{eqnarray}\label{LDPD_step1}
f_{\mathbf{X}_i}  (\mathbf{T}_i \vert
\bolds{\Sigma}, G )=\int \Biggl( N_{2
n} (\mathbf{z}_i\mid\mathbf{X}_i\bolds{\beta},
\bolds{\Sigma}  ) \prod_{j=1}^{2n} T_{ij}^{-1}  \Biggr)\, d G
 (\bolds{\beta}  ),
\end{eqnarray}
where $N_{2n} (\cdot\vert\bolds{\mu}, \bolds{\Sigma}
 )$ refers to a $2n$-dimensional normal distribution with mean
$\bolds{\mu}$ and covariance matrix $\bolds{\Sigma}$. The
mixture model $f_{\mathbf{X}_i}$ can be
equivalently written as a hierarchical model by introducing latent
variables $\bolds{\beta}_i^*$ such that
%
%e2.12 ###
\begin{eqnarray}\label{LDPD_prior.Z}
\mathbf{z}_i \vert\bolds{\beta}^*_i, \bolds{\Sigma}
\stackrel{\mathrm{i.n.d.}}{\sim}
N_{2n} (\mathbf{X}_i\bolds{\beta}_i^*,\bolds{\Sigma} ),
\end{eqnarray}
%
%e2.13 ###
\begin{eqnarray}\label{LDPD_prior.beta}
\bolds{\beta}_1^*,\ldots,\bolds{\beta}^*_m \mid G
\stackrel{\mathrm{i.i.d.}}{\sim} G
\end{eqnarray}
and
%
%e2.14 ###
\begin{eqnarray}\label{LDPD_prior.G}
G\vert a,b,G_0 \sim \mathit{PD}(a,b,G_0),
\end{eqnarray}
where the baseline distribution $G_0$ is assumed to be
$n(p+q)$-dimensional normal distribution
$G_0 (\bolds{\beta} ) =
N_{n(p+q)} (\mathbf{m},\mathbf{S} )$.

%s2.5 ###
\subsection{Some properties}\label{LDPD_sec2.5}
An important property of the proposed model given by expressions (\ref
{LDPD_step1})--(\ref{LDPD_prior.G}) is that the complete distribution
of survival times is allowed to change with values of the predictors
(including properties such as skewness, multimodality, quantiles, etc.)
instead of just one or two characteristics, as implied for many
commonly used survival models. However, we make explicit the dependence
of some functionals of interest of the distribution of the event times
on the covariates in order to compare them to the corresponding
expression arising from the commonly used models. The implied marginal
mean, hazard function and cumulative distribution (CDF) function for
coordinate $j$ in the vector $\mathbf{T}_i$, $T_{ij}$, as functions
of the associated vector of the design matrix $\mathbf{X}_i$,
$\mathbf{x}_{ij}$, are given by
%
%e2.15 ###
\begin{eqnarray}\label{implied_mean}
E(T_{ij}\vert\mathbf{x}_{ij})= \sum_{l=1}^{\infty} \omega_{l} \exp
 \{\mathbf{x}_{ij}'\bolds{\beta}_l +0.5\sigma^2_j  \},
\end{eqnarray}
%
%e2.16 ###
\begin{eqnarray}\label{implied_hazard}
h_{T_{ij}\mid\mathbf{x}_{ij}}(t)=\frac{\sum_{l=1}^{\infty} \omega
_l f_{0,\sigma^2_j}  (\exp \{-\mathbf{x}_{ij}'\bolds{\beta}_l \} t )}{F_{T_{ij}\mid\mathbf{x}_{ij}}(t)}
\end{eqnarray}
and
%
%e2.17 ###
\begin{eqnarray}\label{implied_CDF}
F_{T_{ij}\mid\mathbf{x}_{ij}}(t)=\sum_{l=1}^{\infty} \omega_l
F_{0,\sigma^2_j} (\exp \{-\mathbf{x}_{ij}'\bolds{\beta}_l \} t ),
\end{eqnarray}
respectively, where $f_{0,\sigma^2}$ and $F_{0,\sigma^2}$ refers to the
density and CDF of a lognormal distribution with mean 0 and variance
$\sigma^2$, and $\sigma^2_j=\bolds{\Sigma}_{jj}$. These
expressions show the additional flexibility associated to the proposed
model. For instance, in contrast to a simple AFT survival model based
on the lognormal distribution, the mean function of our proposal given
by expression (\ref{implied_mean}) is a convex combination of
exponential functions. Furthermore, the implied CDF given by expression
(\ref{implied_CDF}) is a convex combination of CDF's arising under the
AFT model,
$F_{T_{ij}\mid\mathbf{x}_{ij}}(t)=F_{0,\sigma^2_j} (\exp \{
-\mathbf{x}_{ij}'\bolds{\beta} \} t )$, where
covariates act multiplicatively on arguments of the baseline survival function.
This simple fact induces an important property of our proposal, namely,
that survival curves are allowed to
cross for different values of a predictor, which is not possible under
the AFT assumption. Other commonly used models such as PH, AH and PO
will also fail to capture this behavior. Under the PH, AH and PO
models, the dependence of the CDF on predictors is given by
\begin{eqnarray*}
1-F_{T_{ij}\mid\mathbf{x}_{ij}}(t)&=& \{1-F_{0,\sigma
^2_j}(t) \}^{\exp \{\mathbf{x}'_{ij}\bolds{\beta} \}},
\\
1-F_{T_{ij}\mid\mathbf{x}_{ij}}(t)&=& \{1-F_{0,\sigma
^2_j}(t) \} \exp \{
-\mathbf{x}'_{ij}\bolds{\beta}
t \}
\end{eqnarray*}
and
\[
\frac{1-F_{T_{ij}\mid\mathbf{x}_{ij}}(t)}{F_{T_{ij}\mid\mathbf{x}_{ij}}(t)}=\frac{1-F_{0,\sigma^2_j}(t)}{F_{0,\sigma^2_j}(t)}\exp
\{\mathbf{x}'\bolds{\beta} \},
\]
respectively. Notice that this constraint associated to the commonly
used models remains if $F_{0,\sigma_j}$ is modeled in a nonparametric
manner and/or if the linear form $\mathbf{x}_{ij}'\bolds{\beta}$ is replaced for a more general function $m(\mathbf{x}_{ij})$.
Although some fixes have been proposed in the context of PH models for
this unappealing property, for example, the inclusion of interactions
with time or stratification, our modeling approach has proved to be a
more flexible alternative. We refer to \citet{deioriorjohnsonmuellerrosnermaceachern2009} for a thorough
comparison in the context of univariate (not doubly censored) survival data.

%s2.6 ###
\subsection{Prior distributions and MCMC implementation}\label{LDPD_sec2.6}
For $a$ and
$b$ we consider joint prior distributions of the kind
$p(a,b)=p(a)p(b\vert a)$, where $p(a)$ is a mixture of point mass at
zero and a continuous distribution on the unit interval $(0,1)$ and
$p(b\vert a)$ is a continuous distribution supported on
$(-a,\infty)$. More specifically, we assume
%
%e2.18 ###
\begin{eqnarray}\label{LDPD_mprior}
a\vert\lambda,\alpha_0,\alpha_1 \sim
\lambda\delta_{0}(\cdot)+(1-\lambda)\operatorname{Beta}(\cdot\vert\alpha_0,\alpha_1)
\end{eqnarray}
and
%
%e2.19 ###
\begin{eqnarray}\label{LDPD_bprior}
b\vert a,\mu_b,\sigma_b \sim N(\mu_b,\sigma_b)I(-a,\infty),
\end{eqnarray}
where $0\leq\lambda\leq1$, and
$\operatorname{Beta}(\cdot\vert\alpha_0,\alpha_1)$ refers to a beta
distribution with parameters $\alpha_0$ and $\alpha_1$. This modeling
strategy allows us to explicitly compare a DP model
versus an encompassing PD alternative. Notice that this is an important
component because the evaluation of any other model comparison criteria
would require the computation of a highly complex area under the
multivariate normal distribution which is difficult to be performed in
practice. Finally, to complete the
model specification, we assume independent hyper-priors
$\mathbf{m} \sim
N_{n(p+q)} (\bolds{\eta},\bolds{\Upsilon} )$,
$\mathbf{S} \sim
IW_{n(p+q)} (\gamma,\bolds{\Gamma} )$, and
$\bolds{\Sigma} \sim
IW_{2n} (\nu,\bolds{\Omega} )$, where
$IW_{2n} (\nu,\bolds{\Omega} )$ denotes a
$2n$-dimensional inverted-Wishart distribution with degrees of
freedom $\nu$ and scale matrix $\bolds{\Omega}$.

The hierarchical representation of the model allows straightforward
posterior inference
with Markov Chain Monte Carlo (MCMC) simulation. As in the context
of standard DP models, two different kinds of MCMC strategies could
be considered for computation in the LDPD model: (I) to marginalize
out the unknown infinite-dimensional distributions [see, e.g.,
\citet{ishwaranjames2003} and
\citet{navarretequintanamuller2008}] or (II) to employ a
truncation to the stick-breaking representation of the process [see,
e.g., \citet{ishwaranjames2001}]. In the case (I), several
alternative algorithms could be considered to sample the cluster
configurations: (I.a) via a Gibbs scheme through the coordinates
[see \citet{navarretequintanamuller2008} for a discussion in
the PD context] or (I.b) to adapt reversible-jump-like algorithms
[see, e.g., \citet{dahl2005}] to the PD context. Functions
implementing these approaches were written in a compiled
language and incorporated into the R library
``DPpackage'' [\citet{jara2007}]. A complete description
of the full conditionals and algorithms is available in the
supplemental article [\citet{jaralesaffredeiorioquintana2010a}].

%s3 ###
\section{An illustration using simulated data}\label{LDPD_sec3}
To validate our approach, we conducted the analysis of real-life and
simulated data sets.
The results of the real-life data analysis are reported in the
supplemental article [\citet{jaralesaffredeiorioquintana2010b}].
The simulated data sets mimic to a certain extent the
Signal-Tandmobiel\tsup{\textregistered} data. We consider one onset time $T_i^O$
and one time-to-event time $T_i^T$ for $m=500$ subjects. We assume a
binary predictor and 250 subjects in each level (groups A and B).
Different distributions were assumed for each level of the predictor
such that
\begin{eqnarray}\nonumber
\log(T_1^O, T_1^T),\ldots,\log(T_{250}^O, T_{250}^T) \vert f_A \stackrel{\mathrm{i.i.d.}}{\sim} f_A
\end{eqnarray}
and
\begin{eqnarray}\nonumber
\log(T_{251}^O, T_{251}^T),\ldots,\log(T_{500}^O, T_{500}^T) \vert f_B
\stackrel{\mathrm{i.i.d.}}{\sim} f_B.
\end{eqnarray}
Two scenarios for the distributional parts of the model were
considered. In scenario
I, a mixture of two bivariate lognormal distributions was assumed for
group A while a bivariate lognormal distribution was assumed for group
B. An important characteristic of scenario I is the bimodal behavior of
the distribution of the onset time and time-to-event in group A. In
group B, a unimodal behavior for the distribution of both variables was
assumed. In scenario II, mixtures of bivariate lognormal distributions
were assumed for both groups. However, the components of the mixtures
were specified in such a way that, for group A, the onset times follow
a bimodal distribution and the time-to-events follow a unimodal
distribution. In group B, the reverse behavior was assumed, namely, the
onset times follow a unimodal distribution while the time-to-events a
bimodal distribution.

In both scenarios and variables of interest, the survival curves for
both groups cross.
The true distributions in each scenario are given next:
\begin{itemize}
\item\textit{Scenario I}: Mixture model for group A--Single model for
group B.
\begin{eqnarray}\nonumber
f_A &\equiv& 0.5 \times N_2
\left(
\left [
\matrix{
1.80 \cr
0.75
}
\right ],
10^{-3}  \left[
\matrix{
5.00 & 2.50\cr
2.50 & 300
}
 \right]
 \right)
\\\nonumber
&&{}+ 0.5 \times N_2  \left(
\left [
\matrix{
2.40 \cr
3.00
}
\right ],
10^{-3}
 \left[
\matrix{
2.50& 1.25\cr
1.25 &100
}
\right ]
 \right)
\end{eqnarray}
and
\begin{eqnarray}\nonumber
f_B \equiv N_2  \left(
\left [
\matrix{
2.1 \cr
2.2
}
\right ],
10^{-2}
\left [
\matrix{
3.24 & 8.10\cr
8.10 & 64
}
\right ]
\right ).
\end{eqnarray}

\item\textit{Scenario II}: Mixture model for both groups A and B.
\begin{eqnarray}\nonumber
f_A &\equiv& 0.5 \times N_2  \left(
\left [
\matrix{
1.8 \cr
2.2
}
\right ],
10^{-3}
\left [
\matrix{
5.50 & 2.50\cr
2.50 & 640
}
\right ]
\right )  \\
\nonumber
&&{}+0.5 \times N_2  \left(
\left [
\matrix{
2.4 \cr
2.2
}
\right ],
10^{-3}
\left [
\matrix{
2.50 & 1.25\cr
1.25 & 640
}
\right ]
\right )
\end{eqnarray}
and
\begin{eqnarray}\nonumber
f_B &\equiv& 0.5 \times N_2  \left(
\left [
\matrix{
2.10 \cr
0.75
}
\right ],
10^{-2}
\left [
\matrix{
3.24 & 8.10\cr
8.10 & 30.00
}
\right ]
\right )
\\\nonumber
&&{}+ 0.5 \times N_2  \left(
\left [
\matrix{
2.10 \cr
0.75
}
\right ],
10^{-3}
\left [
\matrix{
32.4 & 1.25\cr
1.25 & 100
}
\right ]
\right ).
\end{eqnarray}
\end{itemize}
The true onset and event times were interval-censored by simulating the
visit times
for each subject in the data set. The first visit was drawn from an
$N(7,0.2^2)$ distribution. Each of the distances between the
consecutive visits was drawn from an $N(1, 0.05^2)$ distribution.

The LDPD model was fitted to both simulated data sets using the
following values for the hyper-parameters: $\lambda=0.5$, $\alpha
_0=\alpha_1=1$,
$\mu_b=10$, $\sigma_b=200$, $\nu=4$, $\bolds{\Omega}=
\mathbf{I}_2$, $\gamma=5$, $\bolds{\Gamma}=
\mathbf{I}_4$, $\bolds{\eta}=\mathbf{0}_4$ and
$\bolds{\Upsilon}=100 \mathbf{I}_4$. In each analysis 4.02
millions of samples of a Markov chain cycle were completed. Because of
storage limitations and dependence, the full chain was subsampled
every 200 steps after a burn-in period of 20,000 samples, to give a
reduced chain of length 20,000.

\begin{figure}

\includegraphics{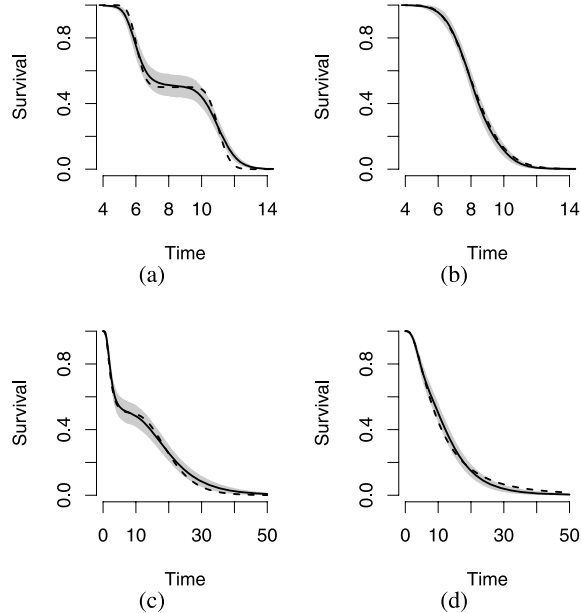}

%{
%% \label{fig1:a}
%}
%{
%% \label{fig1:b}
%}
%{
%% \label{fig1:c}
%}
%{
%% \label{fig1:d}
%}
\caption{\label{LDPD_fig:sim:1} Simulated data---Scenario \textup{1}: Estimated
survival functions for the
onset and time-to-event times for the group A are displayed in panels
\textup{(a)} and \textup{(c)}, respectively.
Estimated survival functions for the
onset and time-to-event times for the group B are displayed in panels
\textup{(b)} and \textup{(d)}, respectively.
The posterior means (solid lines) are presented along the pointwise
95\% HPD intervals. The true functions are presented in dashed lines.}
\end{figure}

\begin{figure}

\includegraphics{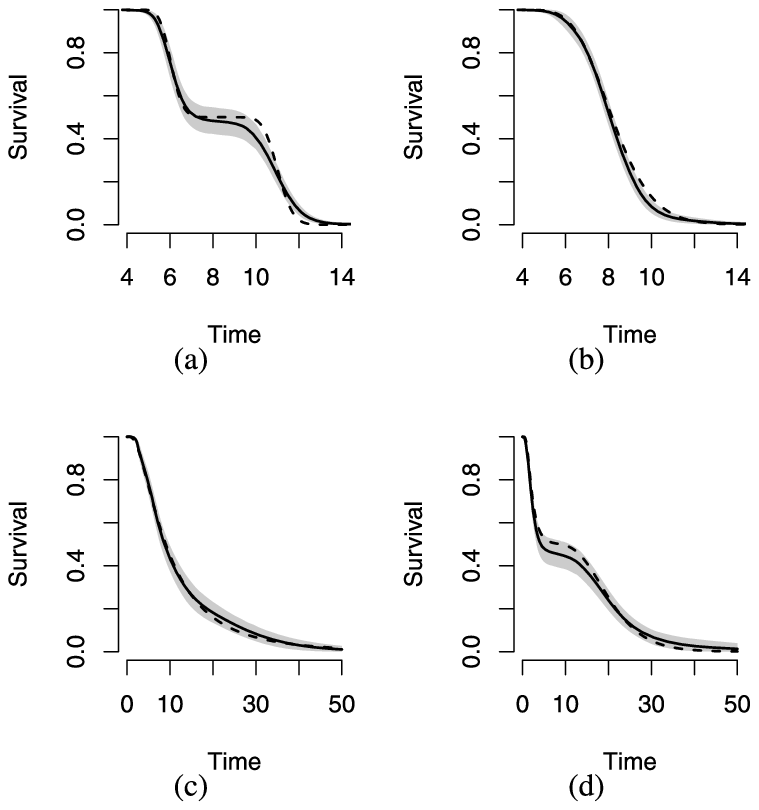}

%{
%% \label{fig1:a}
%}
%{
%% \label{fig1:b}
%}
%{
%% \label{fig1:c}
%}
%{
%% \label{fig1:d}
%}
\caption{\label{LDPD_fig:sim:2} Simulated data---Scenario \textup{2}: Estimated
survival functions for the
onset and time-to-event times for the group A are displayed in panels
\textup{(a)} and \textup{(c)}, respectively.
Estimated survival functions for the
onset and time-to-event times for the group B are displayed in panels
\textup{(b)} and \textup{(d)}, respectively.
The posterior means (solid lines) are presented along the pointwise
95\% HPD intervals. The true functions are presented in dashed lines.}
\end{figure}

Figures~\ref{LDPD_fig:sim:1} and~\ref{LDPD_fig:sim:2} display the true
and estimated survival curves for the onset and time-to-event under
scenarios I and II, respectively. The predictive survival function
closely approximated the true survival functions, which were almost
entirely enclosed in pointwise 95\% highest posterior density (HPD)
intervals. We note that these results are for one random sample from
two particular densities, and these conclusions should not be
overinterpreted. Nonetheless, these examples do show that our proposal
is highly flexible and is able to capture different behaviors of the
onset and time-to-event survival functions. The examples also show that
when a parametric model is appropriated, the proposed model does not
overfit the data.
%
%f3 ###

%f4 ###

%s4 ###
\section{The Signal-Tandmobiel\tsup{\textregistered} data}\label{LDPD_sec4}

%s4.1 ###
\subsection{The Signal-Tandmobiel\tsup{\textregistered} study and the research
questions}\label{sec4.1}

For this project 4468 children were examined on a yearly
basis during their primary school time (between 7 and 12 years of
age) by one of sixteen dental examiners. Sampling of the children
was done according to a cluster-stratified approach with 15 strata.
A stratum consists of a particular combination of one of the five
provinces in Flanders with one of the three school systems. Schools
were selected such that all children had equal probability of being
selected and for each school all children of the first class were
examined. Clinical data were collected by the examiners based on
visual and tactile observations (no X-rays were taken), and data on
oral hygiene and dietary habits were obtained through structured
questionnaires completed by the parents.

The primary interest of our analysis is to study the relationship
between age at start brushing (in years) and deciduous second molars
health status (sound/affected) with caries susceptibility of the
adjacent permanent molars. Here, ``affected molar'' refers to a tooth
that is decayed, filled or missing due to caries. The deciduous second
molars refer to teeth 55, 65, 75 and 85 and first molars refer to teeth
16 and 26 on the maxilla (upper quadrants), and teeth 36 and 46 on the
mandible (lower quadrants). The numbering of the teeth follows the FDI
(Federation Dentaire Internationale) notation which indicates the
position of the tooth in the mouth (see Figure~\ref{fig:teeths}).
Position 26, for instance, means that the tooth is in quadrant 2 (upper
left quadrant) and position 6 where numbering starts from the
mid-sagittal plane. The level of decay was scored in four levels of
lesion severity: $d4$ (dentine
caries with pulpal involvement), $d3$ (limited dentine caries), $d2$
(enamel cavity) and $d1$ (white or brown-spot initial lesions
without cavitation). Here we consider level $d_3$ of severity, which
defines a progressive disease.

Note that for about five years the deciduous second molars are in
the mouth together with the permanent first molars. It is thus
possible that a caries process on the primary and permanent molar
occurs simultaneously. In this case it is difficult to know whether
caries on the deciduous molar caused caries on the permanent molar
or vice versa. For this reason, the permanent first molar was
excluded from the analysis if caries were present when emergence was
recorded. Moreover, the permanent first molar had to be excluded
from the analysis if the adjacent deciduous second molar was not
present in the mouth already at the first examination. For 948 children
none of the permanent first molars was
included in the analysis due to the previously mentioned reasons. In
total, 3520 children (12,485 permanent first molars) were included
in the analysis of which 187 contributed one tooth, 317 two teeth,
400 three teeth and 2616 all four teeth.

%s4.2 ###
\subsection{The analysis and the results}
We consider gender (0${}={}$boy, 1${}={}$girl) and the status of the adjacent
deciduous second molar
(sound${}={}$0, affected${}={}$1) as covariates for the emergence times
$T^O_{ij}$, namely, to define the design vectors $\mathbf{x}^O_{ij}$. For the time-to-caries variables, we use a
similar set of covariates as \citet{leroybogaertslesaffredeclerck2005a}, namely, the covariate vectors
$\mathbf{x}^T_{ij}$ for the caries part of the model include
gender, presence of sealants on the permanent first molar (0${}={}$absent,
1${}={}$present), occlusal plaque accumulation for the permanent first
molar (0${}={}$none, 1${}={}$in pits and fissures or on total surface),
reported oral brushing habits (0${}={}$not daily, 1${}={}$daily) and status of
the adjacent deciduous second molar. In contrast to \citet{leroybogaertslesaffredeclerck2005a}, we did not use the status of
the adjacent deciduous first molar as a covariate due to its large
dependence on the status of the adjacent deciduous second molar and
included the age at start brushing in a linear fashion.

For the model, 4.02 millions of samples of a Markov chain cycle were
completed. Because of storage limitations and dependence, the full
chain was sub-sampled every 200 steps after a burn-in period of 20,000
samples, to give a reduced chain of length 20,000. We consider $\lambda
=0.5$ reflecting equal prior probabilities for
the LDDP and LDPD models. The values of the other hyper-parameters
were taken as $\alpha_0=\alpha_1=1$, $\mu_b=10$, $\sigma_b=200$,
$\nu=10$, $\bolds{\Omega}=\mathbf{I}_8$, $\gamma=31$,
$\bolds{\Gamma}=
\mathbf{I}_{28}$, $\bolds{\eta}=\mathbf{0}_{28}$ and
$\bolds{\Upsilon}=100 \times\mathbf{I}_{28}$. We also
performed the analysis with different hyper-parameters values,
obtaining very similar results. This suggests robustness to the
prior specification.

The posterior probability for $a=0$ was 21.63\%. Correspondingly, the
Bayes factor for the hypothesis of a LDPD against the DP version of
the model was 3.62. This result suggests a ``substantial'' support of
the data to the PD version of the model according to the Jeffreys'
scale [\citet{jeffreys61}, page 432]. As Bayes factors may be
sensitive to the prior specification, we performed a sensitivity
analysis using different prior weights on the LDDP versus a more
general LDPD model. Specifically, we chose $\lambda=0.3$ and
$\lambda=0.7$. The corresponding Bayes factors for the LDPD against
the DP version of the model were 2.72 and 2.21, respectively. The
results, therefore, indicate robustness of the model choice to the
prior specification. More importantly, in all cases the PD version of
the model is to
be preferred when compared to the single precision DP model.

The emergence and caries processes showed a nonsignificant association,
evaluated by the Pearson correlation coefficient on the log-scale
induced by $\bolds{\Sigma}$, for most of the teeth, except for
tooth 46 where a small negative association was observed. The posterior
mean (95\% HPD intervals) for the emergence and caries processes for
tooth 16, 26, 36 and 46
were $-$0.06 ($-$0.18; 0.05), $-$0.06 ($-$0.18; 0.07), $-$0.05 ($-$0.13; 0.02)
and $-$0.10 ($-$0.18; $-$0.02), respectively. The association among
emergence times and among time-to-caries was positive and significant.
Table~\ref{LDPDrho} displays the posterior means and 95\% HPD intervals
for the Pearson correlation among the teeth. The results indicate an
exchangeable correlation matrix would suffice to explain the emergence
process. However, this type of association structure does not hold for
the caries process. The Pearson correlation was bigger for the log
time-to-caries for teeth in the same jaw. Similar and lower
associations were observed when
considering diagonally or vertically opponent teeth. Thus, the results
suggest that the correlation structure induced for frailty models is
not appropriate for these data.
%
%t1 ###
\begin{table}
\caption{Signal-Tandmobiel\tsup{\textregistered} study: Posterior mean (95\% HPD interval) for
the Pearson correlation coefficient between log emergence times (upper
diagonal) and log time-to-caries (lower diagonal) for different teeth}\label{LDPDrho}
\begin{tabular*}{\textwidth}{@{\extracolsep{4in minus 4in}}lcccc@{}}
\hline
& \multicolumn{4}{c@{}}{\textbf{Tooth}}\\[-6pt]
& \multicolumn{4}{c@{}}{\hrulefill}\\
\textbf{Tooth} & \textbf{16} & \textbf{26} & \textbf{36} & \textbf{46}\\
\hline
16 & & 0.60 (0.56; 0.64) & 0.60 (0.56; 0.64)& 0.60 (0.56; 0.64)\\
26 & 0.88 (0.81; 0.94) & & 0.59 (0.55; 0.63) & 0.59 (0.57; 0.63)\\
36 & 0.47 (0.35; 0.57) & 0.43 (0.30; 0.55) & & 0.61 (0.57; 0.65)\\
46 & 0.44 (0.28; 0.61) & 0.39 (0.22; 0.58) & 0.61 (0.54; 0.67) & \\
\hline
\end{tabular*}
\end{table}

In contrast to NPML approaches, an important characteristic of the
proposed model is the ability to make inferences on any quantile of
interest. With respect to the median, neither the emergence nor
the caries process exhibit a significant difference among the four
permanent first molars. For all combinations of covariates,
molars of girls tend to emerge earlier than those of boys. However,
nonsignificant differences were found. Regarding caries
experience, the difference between boys and girls was not significant,
however, the frequency of brushing, presence of sealant, presence of
plaque, age at start brushing and caries experience of neighboring
deciduous second molars have a significant effect on the caries process.
Table~\ref{LDPDmedians} shows the posterior mean and the 95\%
HPD interval for the median emergence time and time-to-caries for
teeth 36 and 46 of boys with the ``best,'' ``worst'' and two
intermediate combinations of discrete covariates. The results are shown
for 4 different values of age at start brushing.
%
%t2 ###
\begin{sidewaystable}
\tablewidth=\textwidth
\centering\caption{Signal-Tandmobiel\tsup{\textregistered} study: Posterior mean
(95\% HPD interval) for
the median emergence time and time-to-caries since emergence
(years) for some covariate combinations and teeth. The results are
shown for boys and teeth 36 and 46 with the following combination of the
covariates: G1 for no plaque, present sealing, daily brushing and sound
primary second molar,
G2 for no plaque, present sealing, daily brushing and affected primary
second~molar, G4
for present
plaque, no sealing, not daily brushing and sound primary second molar, and
G4 for for present
plaque, no sealing, not~daily brushing and
affected primary second molar}\label{LDPDmedians}
\begin{tabular*}{\textwidth}{@{\extracolsep{4in minus 4in}}lcccccll@{}}
\hline
\textbf{Age at start} &  & & \multicolumn{2}{c}{\textbf{Emergence}} & &\multicolumn{2}{c}{\textbf{Caries}}
\\[-6pt]
&& & \multicolumn{2}{c}{\hrulefill} & &\multicolumn{2}{c@{}}{\hrulefill} \\
\textbf{brushing (years)} & \textbf{Covariate group} & & \multicolumn{1}{c}{\textbf{Tooth 36}} & \multicolumn{1}{c}{\textbf{Tooth 46}} & &
\multicolumn{1}{c}{\textbf{Tooth 36}} & \multicolumn{1}{c@{}}{\textbf{Tooth 46}}\\
\hline
1 & G1 & & 6.57 (6.54; 6.60) & 6.56 (6.53; 6.60) &          & 12.62 (11.44; 13.82) & 11.89 (10.65; 13.17)\\
  & G2 & & 6.58 (6.54; 6.61) & 6.57 (6.54; 6.61) & & \phantom{0}9.99 (8.80; 11.18) & \phantom{0}9.72 (8.45; 11.04)\\
  & G3 & & 6.57 (6.54; 6.60) & 6.56 (6.53; 6.60) & & \phantom{0}7.72 (6.68; 8.54)  & \phantom{0}8.49 (6.95; 9.79)\\
  & G4 & & 6.58 (6.54; 6.61) & 6.57 (6.54; 6.61) & & \phantom{0}5.98 (4.98; 6.85)  & \phantom{0}6.83 (5.49; 7.94)\\
[3pt]
3 & G1 & & 6.57 (6.54; 6.60) & 6.56 (6.53; 6.60) & & 11.08 (9.82; 12.29)             & 10.48 (9.24; 11.765)\\
  & G2 & & 6.58 (6.54; 6.61) & 6.57 (6.54; 6.61) & & \phantom{0}8.63 (7.65; 9.73)   & \phantom{0}8.47 (7.23; 9.63)\\
  & G3 & & 6.57 (6.54; 6.60) & 6.56 (6.53; 6.60) & & \phantom{0}6.66 (5.85;  7.46)  & \phantom{0}7.37 (6.32; 8.39)\\
  & G4 & & 6.58 (6.54; 6.61) & 6.57 (6.54; 6.61) & & \phantom{0}5.16 (4.38; 5.94)   & \phantom{0}5.94 (5.04; 6.75)\\
[3pt]
5 & G1 & & 6.57 (6.54; 6.60) & 6.56 (6.53; 6.60) & & \phantom{0}9.67 (8.09; 11.28)  & \phantom{0}9.25 (7.39; 11.29)\\
  & G2 & & 6.58 (6.54; 6.61) & 6.57 (6.54; 6.61) & & \phantom{0}7.49 (6.32; 8.72)   & \phantom{0}7.47 (5.86; 9.18)\\
  & G3 & & 6.57 (6.54; 6.60) & 6.56 (6.53; 6.60) & & \phantom{0}5.78 (4.85; 6.74)   & \phantom{0}6.47 (5.33; 7.65)\\
  & G4 & & 6.58 (6.54; 6.61) & 6.57 (6.54; 6.61) & & \phantom{0}4.47 (3.71; 5.31)   & \phantom{0}5.22 (4.22; 6.20)\\
[3pt]
7 & G1 & & 6.57 (6.54; 6.60) & 6.56 (6.53; 6.60) & & \phantom{0}8.46 (6.50; 10.45)  & \phantom{0}8.28 (5.69; 11.21)\\
  & G2 & & 6.58 (6.54; 6.61) & 6.57 (6.54; 6.61) & & \phantom{0}6.54 (5.07; 8.01)   & \phantom{0}6.69 (4.56; 9.11)\\
  & G3 & & 6.57 (6.54; 6.60) & 6.56 (6.53; 6.60) & & \phantom{0}5.04 (3.91; 6.25)   & \phantom{0}5.76 (4.26; 7.53)\\
  & G4 & & 6.58 (6.54; 6.61) & 6.57 (6.54; 6.61) & & \phantom{0}3.91 (3.00; 4.87)   & \phantom{0}4.65 (3.38; 6.14)\\
\hline
\end{tabular*}
\end{sidewaystable}

%
%f5 ###
\begin{figure}

\includegraphics{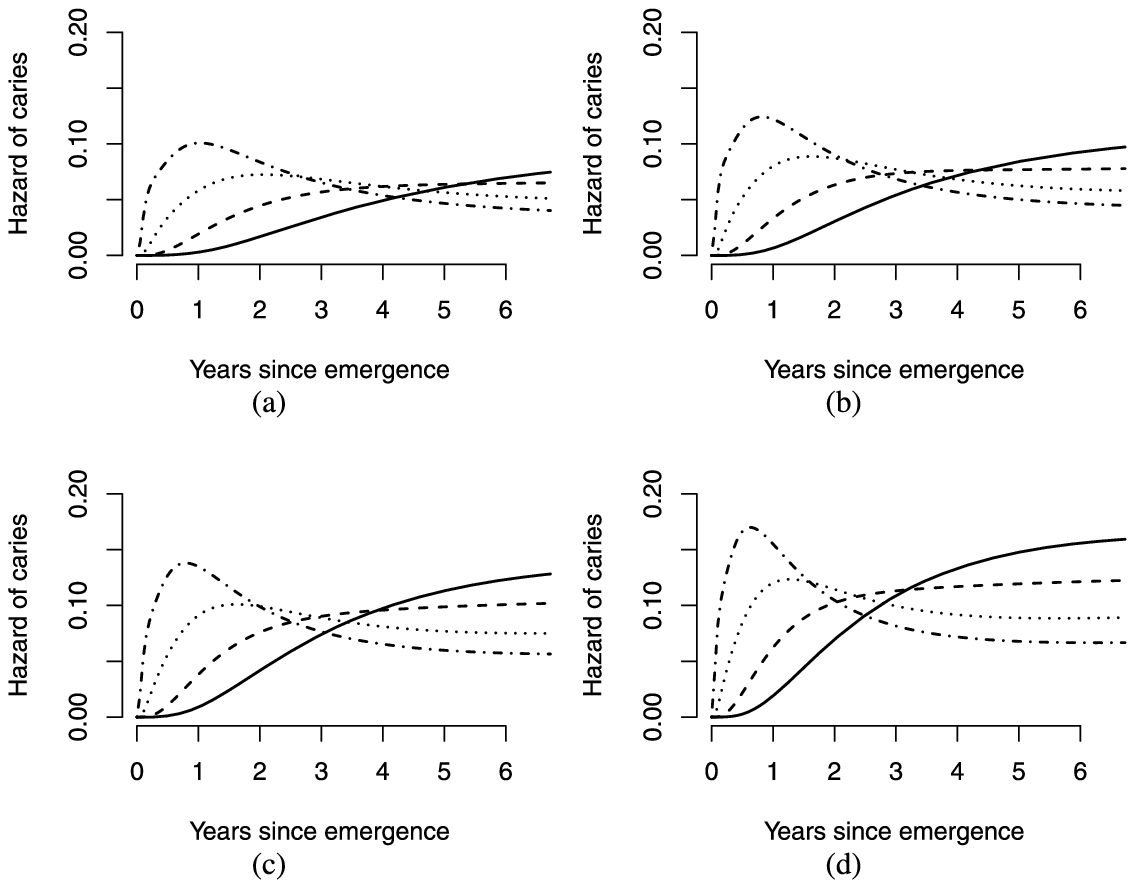}

%[ht]
%{
%% \label{fig1:a}
%}
%{
%% \label{fig1:b}
%}\\
%{
%% \label{fig1:b}
%}
%{
%% \label{fig1:c}
%}
\caption{Signal-Tandmobiel\tsup{\textregistered} study: Estimated hazard function for tooth 16
of boys who started brushing their teeth at the age of 1 (solid line),
3 (dashed line), 5 (dotted line) or 7 (dotted--dashed line). Panels \textup{(a)}
and \textup{(b)} present the results for no plaque, present sealing, daily
brushing and sound primary second molar \textup{(a)}
or affected primary second molar \textup{(b)}. Panels \textup{(c)} and \textup{(d)} present the
results for present
plaque, no sealing, not daily brushing and sound primary second molar
\textup{(c)} or
affected primary second molar \textup{(d)}.}\label{LDPD_fig:st:1}
\end{figure}

\begin{figure}

\includegraphics{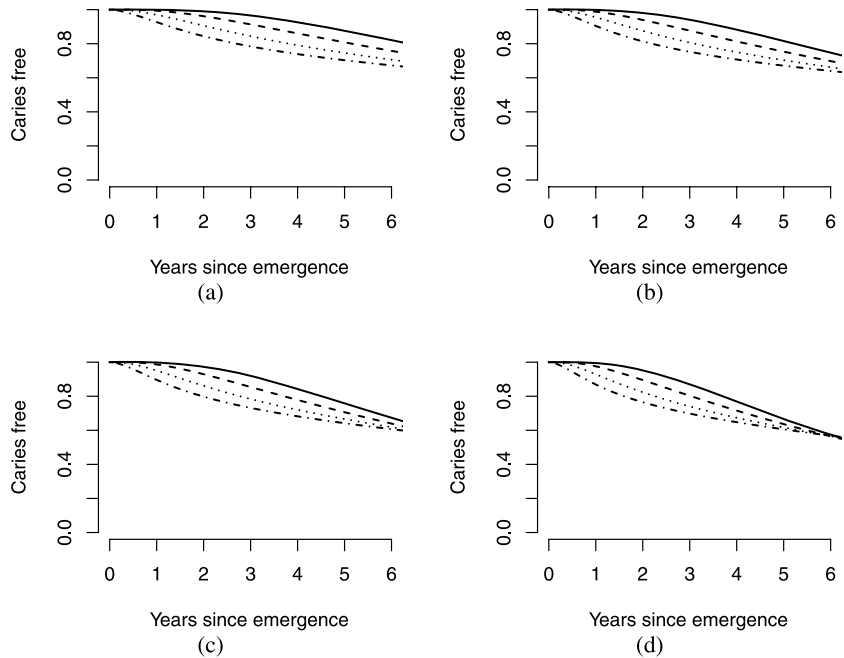}

%[ht]
%{
%% \label{fig1:a}
%}
%{
%% \label{fig1:b}
%}\\
%{
%% \label{fig1:b}
%}
%{
%% \label{fig1:c}
%}
\caption{\label{LDPD_fig:st:2} Signal-Tandmobiel\tsup{\textregistered} study:
Estimated survival function for tooth 16
of boys who started brushing their teeth at the age of 1 (solid line),
3 (dashed line), 5 (dotted line) or 7 (dotted--dashed line). Panels \textup{(a)}
and \textup{(b)} present the results for no plaque, present sealing, daily
brushing and sound primary second molar \textup{(a)}
or affected primary second molar \textup{(b)}. Panels \textup{(c)} and \textup{(d)} present the
results for present
plaque, no sealing, not daily brushing and sound primary second molar
\textup{(c)} or
affected primary second molar \textup{(d)}.}
\end{figure}

Figures~\ref{LDPD_fig:st:1} and~\ref{LDPD_fig:st:2} illustrate the
estimated hazard and survival functions for
the time-to-caries for tooth 16 in boys with the ``best,'' ``worst''
and two intermediate combinations of the discrete covariates by age at
start brushing.
For children who started brushing their teeth after the age of 5, a
high peak in the
hazard function of caries is observed already less than 1 year after
emergence. A smaller peak, shifted to the right and of much lower
magnitude, was observed for children who brush their teeth before the
age of 5. Furthermore, for a given combination of the discrete
predictors, the hazard function for caries crossed for different values
of age at start brushing, suggesting that a proportional hazards model
is not an appropriate alternative for
modeling the time to caries. For a given age at start brushing, the
presence of an affected deciduous second molars significantly increases
the pick in the hazard function of caries in the permanent first molar.
When the teeth are daily brushed since an early age, plaque-free and
sealed the hazard for caries starts to increase approximately 2 years
after emergence, whereas when the teeth are not brushed daily and are
exposed to other risk factors the hazard starts to increase immediately
after emergence. The peak in the hazard for caries after\vadjust{\goodbreak}
emergence can be explained by the fact that teeth are most
vulnerable for caries soon after emergence when the enamel is not yet
fully developed. The curves for girls were similar, and are therefore omitted.

%f6 ###

Figure~\ref{LDPD_fig:st:2} also shows the way in which the age at start
brushing is related to the caries process. The bigger the age at start
brushing, the bigger the prevalence of caries. However, this increase
in the prevalence is only observed in the first years after emergence.
After 5 years since emergence, the prevalence of caries experience
tends to be the same (and can in fact be the same, depending on the
exposure to other risk factors) regardless of the age at start
brushing. This result suggests that PH, AFT, AH or PO models are not
appropriate for the analysis of caries experience since their are
constrained in such a way that survival curves are not allowed to cross
for different values of a predictor. Although the peak in the hazard
for caries at approximately 1--2 years after emergence was also observed
in \citet{leroybogaertslesaffredeclerck2005a} and
\citet{komareklesaffre2008}, this interesting finding was not detected
due to the models considered by these authors.%\looseness=1

%s5 ###
\section{Concluding remarks}\label{discussion}
We have introduced a probability model for dependent random
distributions in the context of multivariate
doubly-interval-censored data. The main features of the proposed
model are ease of interpretation, the ability of testing the hypothesis
of the independence between onset and time-to-event variables,
efficient computation and the fact that assumptions on survival curves,
such as proportional hazards, additive hazards,
proportional odds or accelerated failure time, are not needed.

The proposal is based on a LDPD model, which contains the LDDP model as
an important special case, and is specified in such a way that a simple
hypothesis test for a LDDP versus a more general LDPD alternative can
be performed with no real additional computational effort and without
the need of independent fit of the models.

Several extensions of this work are possible. We are currently working
on a version of the model that takes into account potential
misclassification of the caries process and its effect on the
corresponding inferences. Finally, the extension of the model allowing
for weight dependent covariates is also the subject of ongoing research.

\section*{Acknowledgments}

The first author is supported by the Fondecyt Grant 3095003.
Part of this work was performed when the first and the last two
authors were visiting fellows at the Isaac Newton Institute for
Mathematical Sciences, Cambridge University. The second author has
been supported by the KUL-PUC bilateral (Belgium--Chile) Grant
BIL05/03. The last author has been partially supported by
Fondecyt Grants 1060729 and 1100010, and Laboratorio de An\'alisis
Estoc\'astico PBCT-ACT13.
The authors also acknowledge the partial support from
the Interuniversity Attraction Poles Program P5/24---Belgian State---Federal Office for Scientific, Technical and Cultural Affairs.
Data collection was supported by Unilever, Belgium. The Signal-Tandmobiel\tsup{\textregistered} study comprises the following partners: D.
Declerck (Dental School, Catholic University Leuven), L. Martens
(Dental School, University Ghent), J. Vanobbergen (Dental School,
University Ghent), P.~Bottenberg (Dental School, University Brussels),
E. Lesaffre (Biostatistical
Centre, Catholic University Leuven) and K. Hoppenbrouwers (Youth
Health Department, Catholic University Leuven; Flemish Association
for Youth Health Care).

\begin{supplement}
\sname{Supplement A}\label{suppA}
\stitle{MCMC schemes for posterior computation\\}
\slink[doi]{10.1214/10-AOAS368SUPPA} %,text=10.1214/\break10-AOAS368SUPPA
\slink[url]{http://lib.stat.cmu.edu/aoas/368}
\sdatatype{.pdf}
\sdescription{A complete description
of the full conditionals for marginal and conditional
MCMC algorithms for fitting the LDPD survival model for
doubly-interval-censored data is given.}
\end{supplement}

\begin{supplement}
\sname{Supplement B}\label{suppB}
\stitle{The HIV-AIDS data}
\slink[doi]{10.1214/10-AOAS368SUPPB}
\slink[url]{http://lib.stat.cmu.edu/aoas/368}
\sdatatype{.pdf}
\sdescription{The analysis of the data set considered by \citet{degruttolalagakos89} is presented.
This analysis allows for the comparison of the LDPD model with the
one-sample nonparametric maximum likelihood estimator proposed by \citet{degruttolalagakos89}. The data set considers information from a
cohort of hemophiliacs at risk of human immunodeficiency virus (HIV)
infection from infusions of blood they received periodically to treat
their hemophilia in two hospitals in France. For this cohort both
infection with HIV and the onset of acquired immunodeficiency syndrome
(AIDS) or other clinical symptoms could be subject to censoring.
Therefore, the induction
time between infection and clinical AIDS are treated as doubly-censored.}
\end{supplement}

%suskaldyti doi

% imsref loaded by mstonyte, 2010-09-21 09:54:08

\printaddresses

\end{document}